# On the Existence of Truly Autonomic Computing Systems and the Link with Quantum Computing

Radhakrishnan Srinivasan[1], H. P. Raghunandan[2]
[1]R&D Group, [2]IBM Center for Advanced Studies
*IBM India Software Labs, Golden Enclave, Airport Road, Bangalore 560017, India*
{sradhakr, hpraghu}@in.ibm.com

**Abstract**

*A theoretical model of truly autonomic computing systems (ACS), with infinitely many constraints, is proposed. An argument similar to Turing's for the unsolvability of the halting problem, which is permitted in classical logic, shows that such systems cannot exist. Turing's argument fails in the recently proposed non-Aristotelian finitary logic (NAFL), which permits the existence of ACS. NAFL also justifies quantum superposition and entanglement, which are essential ingredients of quantum algorithms, and resolves the Einstein-Podolsky-Rosen (EPR) paradox in favour of quantum mechanics and non-locality. NAFL requires that the autonomic manager (AM) must be conceptually and architecturally distinct from the managed element, in order for the ACS to exist as a non-self-referential entity. Such a scenario is possible if the AM uses quantum algorithms and is protected from all problems by (unbreakable) quantum encryption, while the managed element remains classical. NAFL supports such a link between autonomic and quantum computing, with the AM existing as a metamathematical entity. NAFL also allows quantum algorithms to access truly random elements and thereby supports non-standard models of quantum (hyper-) computation that permit infinite parallelism.*

## 1. Introduction

The essential requirements of autonomic computing systems (ACS) [1-5] are that they should be self-defining, self-configuring, self-healing, self-optimizing and self-protective. They are also adaptive to their environments, require open industry standards and anticipate the optimized resources needed while keeping their complexity hidden. These requirements, first spelt out by Paul Horn [1], make an ACS an extremely complex object. While the intuitive meanings of these requirements are clear, providing precise mathematical definitions is a very difficult task because of two factors:

(a) In principle, there are infinitely many constraints to be satisfied and this makes an ACS an infinitary entity.
(b) An ACS is seemingly a self-referential entity as well that needs to function, in principle, without human intervention.

In Sec. 2, a simple theoretical model of such an ACS is proposed. Any attempt at a mathematical characterization of an ACS must first necessarily address the issue of existence – can an ACS satisfying such a complex set of requirements exist in principle? In Sec. 3 of this paper, it is demonstrated, via an argument similar to Turing's for the unsolvability of the halting problem, that the existence of a non-trivial ACS is not possible within classical first-order predicate logic (FOPL). The issue of existence is also examined from the point of view of the non-Aristotelian finitary logic (NAFL, see Sec. 5) proposed by the first author [6-8]. It is demonstrated in Sec. 6 that Turing's argument fails in NAFL, which can therefore support the existence of an ACS. Further, NAFL justifies quantum superposition and entanglement, which are essential features of quantum algorithms, and resolves the Einstein-Podolsky-Rosen (EPR) paradox in favour of quantum mechanics (see Sec. 5). It is argued in Sec. 7 that autonomic and quantum computing are possibly linked, *i.e.*, quantum algorithms may be essential for implementing certain aspects of our theoretical model of an ACS. Further, NAFL has the potential to provide the correct logical framework for supporting this link. Therefore further development of NAFL, in particular, an NAFL framework for real analysis and then quantum mechanics, is strongly indicated for theoretical study of quantum and autonomic computing systems. This analysis should also be the starting point for a theoretical description and complexity analysis of practically useful autonomic computing systems, in which the number of constraints is taken as finite. In the concluding remarks (Sec. 8), some brief comments are made on the conceptual advantages of NAFL over intuitionistic, constructive and quantum logics. An overview of our proposed scheme for an ACS is given in Sec. 4.

## 2. Proposed generic model of an ACS

A general computing system (CS) can in principle have infinitely many possible initial states. For example, a network could possibly admit an arbitrary number of servers. Or databases containing an arbitrarily large amount of data can be added to an existing system. For the purposes of this paper, present physical limitations of space or memory, including quantum limitations on miniaturization, are ignored. In general, when a problem -- which could be a virus, for example -- afflicts the CS, it requires a response (automated or through human intervention) from the CS, resulting in a new initial state, which could trigger a new (version of the) problem. This requires another response from the CS and in principle, this loop could continue ad infinitum for the purposes of this paper. In the case of an autonomic system, whether the re-configurations converge to a desirable initial state is an important issue for control theory, but is not dealt with in this paper. Here "problem" is defined very broadly, to include *any* change to the CS, whether deliberate or unanticipated, legitimate or illegitimate. For example, "problems" of a CS could include, among other possibilities, upgradation of computing power and/or memory, partial or total power/hardware failure, virus attacks, server crashes or memory failures leading to loss of data, an increase in the number of users/servers/databases, etc.

The infinite class $S$ of all candidate initial states of a hypothetical computing system and the infinite class $P$ of all candidate "problems" afflicting (all initial states of) this CS are both recursively enumerable, which essentially means that the members of $S$ and $P$ can be enumerated or listed by an algorithm (which may run forever). Of course, such recursive enumerability is a vital requirement for the existence of an ACS; as shown in Sec. 6, this requirement is particularly unproblematic in the logic NAFL [6-8], wherein these classes are recursive. Let $\{S_1, S_2,\ldots, S_i, S_{i+1}, \ldots\}$ and $\{P_1, P_2,\ldots, P_i, P_{i+1},\ldots\}$ be enumerations of $S$ and $P$ respectively. The elements of $S$ and $P$ are essentially classical algorithms that simulate initial states and problems in a suitable formal language. The (ideal, generic) ACS consists of an "autonomic manager" (AM) that monitors the functioning of this CS and takes suitable corrective action whenever problems occur. In particular, AM is equipped with an algorithm $W$ that answers all possible questions of the form "Does initial state $S_i$ correctly handle problem $P_j$ in an optimal manner?". If the answer is "yes", then AM takes no action and if the answer is "no", AM switches CS to a new initial state that does handle $P_j$ in the most optimal manner possible. Of the possibly infinitely many candidates for such a new initial state, AM makes the choice by a suitable optimization. A very important assumption is that the AM is completely hidden from all "problems", in particular, viruses; in other words, AM is not part of any initial state. Otherwise, it is obvious that a virus could be designed to disrupt the functioning of AM and make the purported ACS fail (i.e., require human intervention). Clearly, this assumption requires that the software for AM must be encrypted in such a way that it is *in principle* unbreakable. This is not possible with the classical encryption techniques permitted by FOPL, but can be realized if quantum encryption becomes feasible (see the sections on classical and quantum cryptography in Ref. [26]; in particular, note that the unbreakable codes claimed in classical cryptography require the generation of truly random numbers, not possible in classical computation). As will be shown in Sec. 7, this is one of the links between autonomic and quantum computing that can be supported in NAFL. The above definition of an ACS could, in principle, incorporate all the requirements mentioned in Sec. 1.

## 3. ACS cannot exist within FOPL

The reader may recognize the similarity to Turing's halting problem when considering the question of existence of the algorithm $W$. A diagonalization argument can be used as follows, to show that $W$ cannot exist according to FOPL (which permits only standard computing techniques or Turing machines). Let $V$ be a problem that uses $W$ to function as follows. If $W$ computes that the $i^{th}$ problem $P_i$ is not correctly handled by the $i^{th}$ initial state $S_i$, then $V$ decides not to affect $S_i$; on the other hand, if $W$ computes that $P_i$ is correctly handled by (or, in general, does not affect) $S_i$, then $V$ decides to defeat any protective measures of $S_i$ (if they exist) and affect $S_i$. Since $V$ must belong to any listing of $P$, we obtain the usual contradiction that some initial state must both be affected by and not affected by $V$. From this contradiction, one concludes that the list of yes-no answers required to be provided by $W$ is not recursively enumerable in FOPL and therefore $W$ does not exist. Hence FOPL predicts that genuinely autonomic computing systems that use only classical Turing machines cannot exist.

Note that $V$ may be thought of as a master virus, while AM includes within it a master anti-virus system. Here we have assumed that given any anti-virus system, there always exists a virus that can fool it and that $V$ will be able to generate the appropriate virus software. This assumption is justified for the purposes of establishing the proof-by-contradiction within FOPL, since if $W$ exists as assumed, it must already be able to analyze the anti-virus software of an arbitrary initial state to decide if it can correctly handle an arbitrary virus. One can easily see that $V$ has essentially the same capability as AM, which has already been assumed to exist for the purposes of the proof. The assumption that AM is hidden from all problems (in particular, *V*) cannot be enforced within FOPL, as mentioned in Sec. 2. This failure, which causes the above contradiction, can be circumvented in the logic NAFL [6-8], as noted in Secs. 6-7. Tien Kieu [9-12] has argued that quantum algorithms can in fact sidestep Cantor's diagonalization argument (used in Turing's proof) and therefore the quantum version of *W* can, in principle, exist. In what follows, we suggest that Kieu's thesis can possibly be supported in NAFL, which rejects Cantor's diagonalization argument (see Sec. 6) and has the potential to admit both autonomic and quantum computing systems.

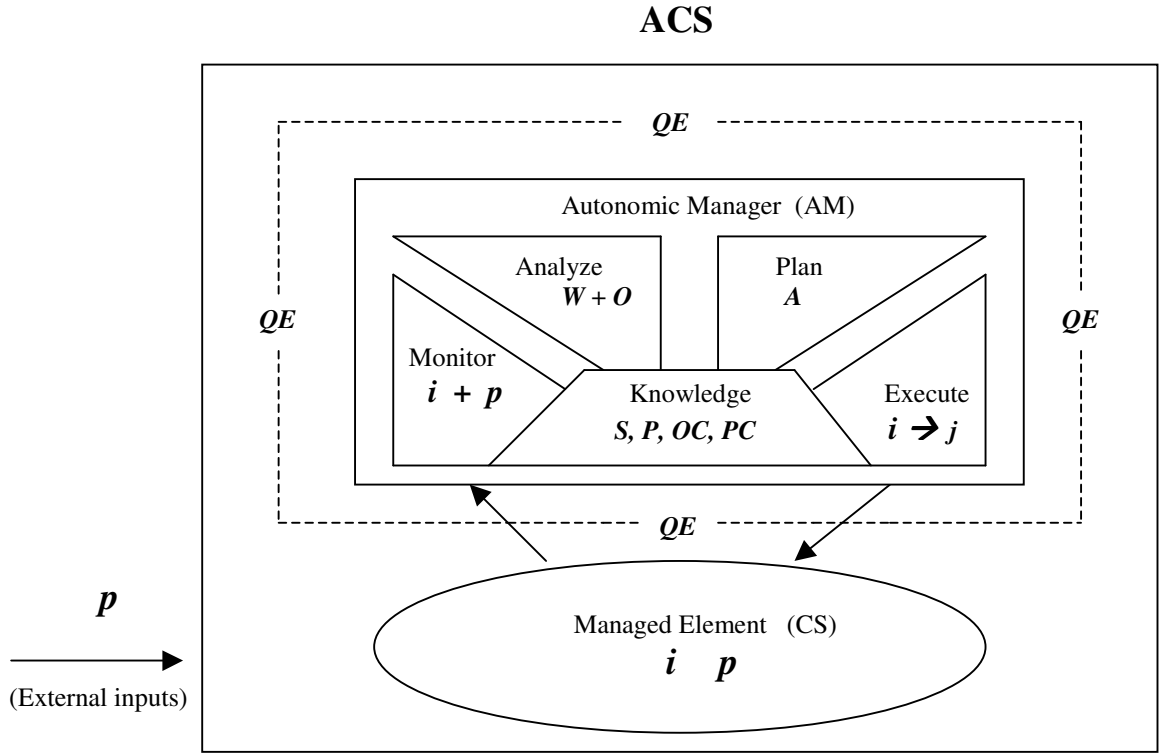

Figure 1. Structure of the proposed autonomic computing system

## 4. Summary of proposed scheme for an ACS

Fig. 1 illustrates our proposed scheme. The notation is as follows.

| | |
|---|---|
| ACS | Autonomic computing system (autonomic element). |
| AM | Autonomic manager. |
| CS | Computing system (managed element). |
| *i, j* | Initial states of CS. |
| *p* | Problem associated with CS. |
| QE | Quantum encryption software. |
| S | The infinite class of all initial states (generated algorithmically). |
| P | The infinite class of all problems (generated algorithmically). |
| OC | Optimization criteria |
| PC | Planning criteria |
| W | Algorithm to determine if an arbitrary *p* is correctly handled by an arbitrary *i*. |
| O | Optimization software (used by *W*) that determines candidate initial state(s) *j* to handle *p* optimally. |
| A | Auxiliary software used to formulate a plan of action based on the analysis returned by the Analyze module and possibly other factors. |

Here *A* determines the unique initial state *j* to handle a given problem *p* and schedules any required switch from *i* to *j*. Note that *A* may possibly have to decide between multiple candidates for *j* returned by the Analyze module, and may also overrule all candidates in special cases. Fig. 1 builds upon the architecture in Ref. [3] (see pp. 43-45 and Fig. 2 therein). The new features of our proposed scheme are as follows.

- Our ACS is meant to be truly autonomic, handling infinitely many possible problems and initial states

of the CS. The external inputs to the CS from other ACS or humans are also classified as problems of the CS, as illustrated in Fig. 1.

- The algorithms used by AM are quantum algorithms (whose metamathematical existence is supported by NAFL) in our scheme. Here AM is an entity that is distinct from the CS, which uses only classical computing techniques; in particular, *S* and *P* are classical entities. Further, AM is insulated from the CS and all problems by in-principle-unbreakable quantum encryption (as shown in Fig. 1 by the dashed lines), so as to practically enforce the above-noted distinction between the AM and the CS and ensure that the ACS does indeed remain fully autonomic.

- The above is in contrast to the scheme proposed in Ref. [3], wherein the authors state on pg. 44 that "*Fully autonomic computing is likely to evolve as designers gradually add increasingly sophisticated autonomic managers to existing managed elements. Ultimately, the distinction between the autonomic manager and the managed element may become merely conceptual rather than architectural, or it may melt away – leaving fully integrated, autonomic elements with well-defined behaviours and interfaces, but also with few constraints on their internal structure.*" Our analysis shows that the existence of such a self-referential entity as proposed by these authors is not logically possible in the fully autonomic case (by Turing's argument in classical logic and by the Main Postulate in NAFL; see Sec. 3 and the ensuing sections). The distinction between the AM and the CS must necessarily be maintained, say, by using quantum algorithms and quantum encryption as we have proposed.

Other than these distinctions, the modules in Fig. 1 titled Monitor, Analyze, Plan and Execute function largely as envisaged by the authors of Ref. [3], drawing upon a common Knowledge element.

## 5. Introduction to NAFL

In this section, we give, for the sake of completeness, a brief description of NAFL and explain how it justifies quantum superposition and entanglement (which are vital ingredients of quantum algorithms). In particular, NAFL resolves the Einstein-Podolsky-Rosen (EPR) paradox in favour of quantum mechanics; see Sec. 5.2. This description will help the reader appreciate the arguments given in the ensuing sections on the importance and potential of NAFL for the theoretical understanding of autonomic/quantum computing systems. See Refs. [6-8] for more details on NAFL.

In NAFL, there are no truths in just the *language* of a theory, unlike classical/intuitionistic/constructive logics – truths for formal propositions exist only with respect to axiomatic theories, which use the same language, well-formed formulae and rules of inference as FOPL (for convenience, we restrict ourselves to FOPL systems which use natural deduction, although this is not essential). However, the Main Postulate of NAFL, explained below, imposes restrictions on NAFL theories. There do exist absolute (Platonic, metamathematical) truths in NAFL, but these are truths *about* axiomatic theories and their models. As in FOPL, an NAFL theory T is defined by its axioms and the theorems deduced from the axioms, using the inference rules; T is defined to be *consistent* if and only if there exists a model for T and a proposition *Q* is *undecidable* in T if and only if neither *Q* nor its negation ¬*Q* is provable in T.

### 5.1. The Main Postulate of NAFL and justification of quantum superposition

If a proposition *Q* is provable/refutable in a consistent NAFL theory T, then it is true/false with respect to T (henceforth abbreviated as "true/false in T"), *i.e.*, a model for T will assign *Q* to be true/false. If *Q* is undecidable in a consistent NAFL theory T, then the Main Postulate of NAFL asserts that *Q* is true/false in T if and only if *Q* is provable/refutable in an *interpretation* T* of T. Here T* is an axiomatic NAFL theory that, like T, resides in the human mind and acts as the "truth-maker" for (a model of) T. Note that for a fixed T, T* could vary in time according to the free will of the human mind that interprets T; for example, T* could be T+*Q* or T+¬*Q* or just T itself at different times for a given human mind, or in the context of quantum mechanics, for a given 'observer'. Further, T* could vary from one observer to another at any given time; each observer determines T* by his or her own free will. The essential content of the Main Postulate is that a proposition *Q* that is undecidable in a consistent NAFL theory T is true/false in T if and only if it has been *axiomatically* declared as true/false with respect to T (by virtue of its provability/refutability in T*). In the absence of any such axiomatic declarations, i.e., if T* is chosen such that *Q* is undecidable in T*, then *Q* is neither true nor false in T, and consistency of T demands that the law of the excluded middle and the law of non-contradiction fail in a non-classical model for T in which *Q*&¬*Q* is the case (see Proposition 1 of Ref. [6] and its proof, which is reproduced here in Appendix A). In this non-classical model, '*Q*' denotes that "¬*Q* is not provable in T*" and '¬*Q*' denotes that "*Q* is not provable in T*". Thus one can see that *Q*, ¬*Q* and *Q*&¬*Q* are indeed *non-classically* true in this model, and the quantum superposition principle is justified by identifying

"measurement" in the real world with these "axiomatic declarations" of truth/falsity of $Q$ (by virtue of its provability/refutability in T*). However, this identification is only an informal convention and need not always hold, as explained in Sec. 5.2. Note also that $Q$ and $\neg Q$ are *classically* 'neither true nor false' in the non-classical model, where 'true' and 'false' have the (classical) meanings given in the Main Postulate.

See Sec. 2 of Ref. [6] for a detailed discussion of the Schrödinger cat example in NAFL. The essence here is that an observer who 'measures' the cat to be alive or dead makes the corresponding axiomatic declaration via its provability in the interpretation QM* of a suitably formalized theory QM of quantum mechanics. Thus when the cat is measured to be alive (dead), the observer could take QM* = QM + $Q$ ($\neg Q$), where $Q$ is the QM-undecidable proposition "The cat is alive". Until such an axiomatic declaration is made (which need not coincide in time precisely with the measurement), the cat is in a superposed state of "neither alive nor dead", *i.e.*, in the state $Q\&\neg Q$; this superposed state means that the cat has not been axiomatically asserted (and in tune with the above convention, measured) by the observer to be either alive or dead. It is important to note that the axiomatic declarations are made by the observer using his/her free will; an NAFL theory T only "sees" these axiomatic declarations of truth which may, but need not, coincide with what the observer sees in the real world. Of course, if the proposition is not about the real world, then no "measurement" exists. NAFL correctly handles the temporal nature of truth via the time dependence of the interpretation T* of a theory T; if the cat is put into the box at $t = 0$ and measured (axiomatically asserted) as alive at $t = 1$, then the proposition $R$ that "The cat was alive for $0 < t < 1$" can be formalized and proven in the interpretation QM* because $R$ would only apply for $t \geq 1$. Note that $R$ does not conflict with the superposed state $Q\&\neg Q$, which applies for $0 < t < 1$.

NAFL is more in tune with the Copenhagen interpretation (CI) of quantum mechanics than the many-worlds interpretation (MWI). Bohr's principle of complementarity easily follows in NAFL. To assert that one cannot measure both $Q$ and $\neg Q$ in a single experiment (where $Q$ is undecidable in QM), translates in NAFL to the requirement that $Q\&\neg Q$ cannot be asserted as an axiom of QM*. Indeed, $Q\&\neg Q$ is not even a legitimate proposition of QM (see the ensuing paragraph) and such an illegal axiomatic assertion would clearly make QM* inconsistent. Here it should be noted that NAFL is not a paraconsistent logic [25]; see the last paragraph of Sec. 7 for further discussion of this fact. But NAFL differs from CI in that it does not require the "collapse of the wavefunction" (see Sec. 5.2). The non-classical model in which quantum superposition holds is a superposition of two or more classical models (or 'worlds'); here '(non)-classical' is used strictly with respect to the status of the proposition $Q$ which is being considered. In this limited sense, MWI is vindicated in NAFL. NAFL is the only logic that correctly embodies the philosophy of formalism [8]; NAFL truths for formal propositions are axiomatic, mental constructs with strictly no Platonic world required. As noted earlier, truths *about* axiomatic NAFL theories are Platonic and the law of the excluded middle applies to such truths. Thus an NAFL theory is either consistent or inconsistent and a formal proposition is either provable or refutable or undecidable in that theory, irrespective of our present state of knowledge concerning these truths. These Platonic notions, which include that of provability, undecidability and the existence of a (non-classical) model for an NAFL theory (and hence, quantum superposition and entanglement) are strictly metamathematical concepts that are not formalizable in NAFL theories.

An NAFL theory T requires two levels of syntax [6, 8], namely, the 'theory syntax' and the 'proof syntax'. The theory syntax consists of precisely those propositions that are legitimate, *i.e.*, whose truth in T satisfies the Main Postulate; obviously, the axioms and theorems of T are required to be in the theory syntax. Further, one can only add as axioms to T those propositions that are in the theory syntax. In particular, neither $Q\&\neg Q$ nor its negation $Q\vee\neg Q$ is in the theory syntax when $Q$ is undecidable in T; this will be justified below. The proof syntax, however, is classical because NAFL has the same rules of inference as FOPL; thus $\neg(Q\&\neg Q)$ is a valid deduction in the proof syntax and may be used to prove theorems of T. For example, if one is able to deduce $A\Rightarrow Q\&\neg Q$ in the proof syntax of T, where $Q$ is undecidable in T and $A$ is in the theory syntax, then one has proved $\neg A$ in T despite that fact that $\neg(Q\&\neg Q)$ is not a theorem (in fact, not even a legitimate proposition) of T. This is justified as follows: $\neg(Q\&\neg Q)$ may be needed to prove theorems of T, but it does not follow in NAFL that the theorems of T imply $\neg(Q\&\neg Q)$ when $Q$ is undecidable in T. Let $A$ and $B$ be undecidable propositions in the theory syntax of T. Then $A\Rightarrow B$ (equivalently, $\neg A\vee B$) is in the theory syntax of T if and only if $A\Rightarrow B$ is *not* (classically) deducible in the proof syntax of T. It is easy to check that if $A\Rightarrow B$ is deducible in the proof syntax of T, then its (illegal) presence in the theory syntax would force it to be a theorem of T, which is not permitted by the Main Postulate: in a non-classical model for T in which both $A$ and $B$ are in the superposed state, $A\&\neg B$ must be non-classically true. If one replaces $B$ by $A$ in this result, one obtains the previous conclusion that $\neg(A\&\neg A)$ is not in the theory syntax. For example, take $T_0$ to be the null set of axioms. Then nothing is provable in $T_0$, *i.e.*, every legitimate proposition of $T_0$ is undecidable in $T_0$. In particular, the proposition ($A$&($A$⇒$B$))⇒$B$, which is deducible in the proof syntax of $T_0$ (via the *modus ponens* inference rule), is not in the theory syntax; however, if $A$⇒$B$ is not deducible in the proof syntax of $T_0$, then it is in the theory syntax. Note also that ¬¬$A$⇔$A$ is not in the theory syntax of $T_0$; nevertheless, the 'equivalence' between ¬¬$A$ and $A$ holds [8] in the sense that one can be replaced by the other in every model of $T_0$, and hence in all NAFL theories. Indeed, in a non-classical model for $T_0$, this equivalence holds in a non-classical sense [8] and must be expressed by a different notation.

The notion of 'observer' and 'measurement' in the real world or that of 'axiomatic declaration' of truth cannot be formalized in the theory syntax of NAFL theories (see Sec. 2 of Ref. [6]). To see this, again consider the Schrödinger cat example and assume to the contrary that $A$ is the formalized version of the proposition that "At $t = 1$ the observer measured (axiomatically declared) the cat to be alive". Then QM does not prove either $A$ or ¬$A$, but would require $A$∨¬$A$ to be a theorem, in violation of the Main Postulate (see Proposition 1 of Appendix A). Hence $A$ cannot be in the theory syntax of QM; however, $A$∨¬$A$ is a legitimate deduction in the proof syntax of QM.

## 5.2. Quantum entanglement explained and EPR paradox resolved in NAFL

Let $A$ ($B$) be an undecidable proposition of QM about a given particle X (its distant entangled counterpart Y) such that $A$⇔$B$ is deducible in the proof syntax of QM. If no measurements are made, let the observer set QM* = QM, so that both $A$ and $B$ are in the superposed state. Note that $A$&¬$B$ (or $B$&¬$A$) is non-classically true in the resulting non-classical model of QM, which explains why $A$⇔$B$ cannot be a theorem of QM in NAFL and is therefore not in its theory syntax. If $A$ is measured at a given time, the X-observer (who is local to the particle X and learns of this measurement) sets QM* = QM + $A$ and it follows that QM* must prove $B$ in NAFL. Thus the axiomatic declaration of $A$ via QM* entails the simultaneous axiomatic declaration of $B$ (with respect to QM) and there is no mystery associated with entanglement in NAFL. It is only when $A$ is interpreted strictly as a 'measurement' on the particle X, as is necessary in standard QM, that one is at a loss to explain how a simultaneous measurement happened on its distant entangled counterpart Y. Indeed an observer who has not learned of the measurement of $A$, such as, the Y-observer (local to the particle Y), will not make any axiomatic declarations and will therefore continue to have QM* = QM in *his* interpretation. Thus both $A$ and $B$ continue to be in the superposed state for the Y-observer, despite the axiomatic declarations of the X-observer. There is no contradiction involved here, because the superposed state only means that the Y-observer has made no axiomatic declarations regarding $A$ and $B$ in *his* interpretation. Indeed, the Y-observer could even axiomatically declare $A$ and $B$ to be *false*, using his free will, despite that fact that he will always measure $B$ to be true (given that the X-observer has measured $A$ to be true). As noted earlier, it is each observer's free will that determines his or her axiomatic declarations (or their absence) and there is no inconsistency if different observers have different, even contradictory, interpretations of QM in mind.

Indeed, the fact that $B$ *could* be made false for the Y-observer via an axiomatic declaration, despite the X-observer's measurement (and axiomatic declaration) of $A$, is crucial for the justification of the superposed state of $B$ and $A$ from the Y-observer's point of view. This would not be possible if $B$ is interpreted strictly as a measurement, for as noted earlier, there is no possibility for the Y-observer to measure $B$ to be false, given the X-observer's measurement. It is in this sense that the earlier proposed (informal) convention of identifying 'measurement' with 'axiomatic declaration' breaks down. Hence standard QM (unlike NAFL) would need the infamous "instantaneous collapse of the wavefunction" to argue that the X-observer's measurement of $A$ instantaneously destroys the superposed state of $B$ for the Y-observer. In NAFL, if and when the Y-observer learns of the X-observer's measurement, or when the Y-observer makes a measurement on Y, he could set QM* = QM + $B$ and deduce $A$ in QM*. Note that both QM + $A$ and QM + $B$ are identical theories that effectively declare A and $B$ axiomatically with respect to QM. The Y-observer, *after* his measurement (and axiomatic declaration) of $B$, could also assert retroactively that $A$ and $B$ *were* true from the (earlier) instant at which the X-observer made his measurement of $A$; as noted in the context of the Schrödinger cat example in Sec. 5, such a retroactive assertion would not temporally contradict the superposed state of $A$ and $B$ which actually held for the Y-observer at that instant (and later).

Non-locality is not a problem in NAFL, which rejects the relativity theories and non-Euclidean geometries [13-16] for essentially the same reason that it accepts superposition and entanglement – the Main Postulate. In Refs. [14-16], propositions *P* and *Q* (which could be based on probabilistic events, for example) have been exhibited such that the Lorentz transformations require *P*⇔*Q* to be a theorem of the theory of special relativity (SR). It immediately follows that the undecidability of *P* and *Q* in SR would make SR inconsistent in NAFL (it is argued in Refs. [14-16] that such inconsistency can be deduced even within FOPL). Essentially the same argument can be adapted to the present context of entanglement, if one treats *A* and *B* as undecidable propositions describing probabilistic spacetime events regarding the distant entangled particles X and Y (see Appendix B). It can be shown that the Lorentz transformations would again require $A \Leftrightarrow B$ to be a theorem of QM despite the undecidability of *A* and *B* in QM; such theoremhood is prohibited in NAFL as noted earlier. Thus NAFL, which requires space to be Euclidean and time to be absolute, rejects the presently accepted arguments that quantum entanglement is compatible with the relativity theories despite non-locality. These arguments usually assert that the said compatibility follows from the fact that there is no faster-than-light *information* transfer from X to Y, despite the presence of non-local correlations in the properties of these particles. In summary, NAFL resolves the celebrated Einstein-Podolsky-Rosen (EPR) paradox [17] in favour of quantum mechanics and non-locality, as opposed to relativity and locality.

It turns out that the requirement of the existence of the non-classical model, which is the main difference between NAFL and FOPL, severely restricts classical infinitary reasoning. For the purposes of this paper, the most important restriction is that infinite sets cannot exist in consistent NAFL theories, essentially because they are self-referential objects whose undecidability of existence in a theory postulating only hereditarily finite sets would violate the Main Postulate (see Sec. 3 of Ref. [7]). However, an infinite proper class can and indeed, *must* exist within an NAFL theory whenever that theory admits infinitely many objects described by that class, which is identified by a suitable property in the language of the theory. For example, if a theory, such as, Peano Arithmetic, admits infinitely many natural numbers, then that theory must necessarily prove [8] the existence of the infinite proper class of all natural numbers. This is a vital difference between NAFL and FOPL that causes these two logics to diverge profoundly in their predictions. In FOPL theories, infinitely many objects satisfying a given property, such as, natural numbers, can exist without entailing the existence of either the infinite class or infinite set of these objects within that theory – such existence requires an additional postulate. Consequently, Peano Arithmetic admits non-standard models in FOPL, but not in NAFL, where the existence of non-standard integers would be a contradiction [8]. Crucially, quantification over infinitely many infinite (proper) classes is banned in NAFL because such quantification treats infinite classes as mathematical objects, *i.e.*, as sets. Cantor's diagonalization argument fails in NAFL, essentially because it is self-referential in nature and requires illegal quantification over proper classes (e.g., real numbers). For the same reasons, Gödel's incompleteness theorems do not apply to, and Turing's halting problem is decidable in, consistent NAFL theories [6, 8]. Hence consistency of Peano Arithmetic demands its *completeness* in NAFL, as opposed to incompleteness in FOPL.

## 6. The failure of Turing's argument in NAFL

Let us examine further, from the point of view of NAFL, the reasons for the decidability of the halting problem. A Turing machine, *by definition*, must either halt or not halt. Hence if *Q* is the proposition that a given Turing machine halts, then $Q \vee \neg Q$ is unavoidably built into the definition of that Turing machine. It then follows that *Q* cannot be undecidable in any consistent NAFL theory in which the existence of that Turing machine is formalized – the required non-classical model for that theory in which $Q \& \neg Q$ is the case cannot exist as demanded by the Main Postulate of NAFL because the said theory *proves* $Q \vee \neg Q$ (see Proposition 1 of Appendix A). In fact, *any* infinite proper class must be recursive (and hence, recursively enumerable) in NAFL theories – whether a given object belongs to that class cannot be undecidable in a consistent NAFL theory because the required non-classical model for that theory in which the said object neither belongs to nor does not belong to that class cannot exist. Such existence would violate the axiom of extensionality for classes, which is an essential ingredient of any consistent NAFL theory in which classes exist. Hence there must necessarily exist an algorithm (but not necessarily a *classical* algorithm, as explained in Sec. 7) for enumerating the elements of any infinite proper class in NAFL, and in particular, Turing's halting routine *H* must exist in NAFL, unlike FOPL.

Another way of understanding the argument for the existence of *H* in NAFL is as follows. Let *C* be the class of all program-data set pairs, i.e., *C* consists of all elements of the form $\{P_i, D_j\}$ where $P_i$ is an arbitrary program (which is an instruction set for a Turing Machine) and $D_j$ is an arbitrary data input. For any set consisting of an arbitrary finite number *k* of elements of *C*, there does exist in NAFL a halting routine $H_k$ which correctly outputs all the yes-no decisions on whether a program in a given element of that set halts or not halts on the data input in that element. But then in NAFL, it necessarily follows that the corresponding infinite class of halting routines $\{H_k\}$ must also exist, or in other words, there must exist a property in the language of the theory (in which the halting problem is formalized) that defines this infinite class. Since non-standard models for arithmetic do not exist in NAFL [8], this infinite class must be recursive. Indeed, the assertion that $H_k$ exists for *arbitrary* (finite) *k* already uses the fact that *k* can only be *standard finite* in NAFL, *i.e.*, non-standard integers do not exist, and amounts to requiring that there exists a bijection between the class *N* of all natural numbers and the class $\{H_k\}$. The existence of such a mapping in NAFL means that there exists an algorithm that will identify the $k^{\text{th}}$ element of the class $\{H_k\}$ for an *arbitrary* $k$, but such an algorithm could be non-classical and hence need not necessarily exist within the theory that proves the existence of $\{H_k\}$, as will be elaborated upon in Sec. 7. The required halting routine $H$ can then be constructed upon using this mapping; $H$, which cannot exist in FOPL, will give the correct halting decisions for all the infinitely many elements of $C$. The intuitive reasoning here is as follows; the existence of infinitely many algorithms in the class $\{H_k\}$ means that 'all' the halting decisions associated with the class $C$ have been listed algorithmically, by the elements of $\{H_k\}$. In NAFL, this is only possible if there *also* exists *an algorithm $H$* that lists the class of infinitely many halting decisions associated with $C$. Unlike in classical logic, an infinite class of NAFL cannot be "completely" listed by the finitary process of counting one (or finitely many) elements at a time [15]; indeed, by induction, such a counting process can never be considered complete, for there will always exist infinitely many elements remaining to be counted. Hence the existence of the class $\{H_k\}$ implies the existence of $H$ in NAFL, although $H$ need only exist metamathematically, as will be explained in Sec. 7.

In the NAFL model of computation, an algorithm (even if it has a classical representation) computes an infinite class by accessing, at some point, a *truly random* element of that class. In NAFL, the Main Postulate requires that such a random or arbitrary element, when its value is not specified, is in a superposed (or universally quantified) state of assuming all possible values in the class [8, 15] *at the same time*. Thus NAFL supports the non-standard models of quantum (hyper-)computation that allow for infinite parallelism [27], and also supports Kieu's claim [9] that hyper-computation is possible because of the ability of quantum algorithms to compute truly random numbers. Such hyper-computation is ruled out in the standard models of quantum computation [26].

Turing's construction of a program that calls $H$ to implement the anti-diagonal halting decisions is banned in NAFL because the said program has to necessarily be impredicatively defined using the class of 'all' programs and is self-referential in nature. Indeed, to even speak of the diagonal sequence is illegal in NAFL because it can only be defined as a diagonal by illegal quantification over proper classes. Careful consideration of Turing's argument will show that it treats a program together with its associated infinitely many possible data inputs as a 'completed' infinite mathematical entity and quantifies over infinitely many such entities, effectively treating these as infinite sets. Such illegal (from the point of view of NAFL) quantification over proper classes is present in *every* version of Turing's argument for the unsolvability of the halting problem (or for the existence of non-computable functions) and in Gödel's incompleteness theorems, all of which make use of some form of Cantor's diagonalization. Here one must note that in some versions of Turing's argument which make use of explicit self-reference, the said quantification is seemingly masked, but is nevertheless present when one takes into account the elaborate procedure (as established by Gödel's diagonalization lemma and the Gödel-numbering scheme) needed to encode the explicit self-reference. As an example, consider the version of Turing's argument where a program $X(x)$ (with data input $x$ also set as equal to $X$) calls the halting routine $H$ to determine whether $X(X)$ halts and then, to establish the contradiction, does the opposite of whatever $H$ determines. The process of encoding the self-reference requires the diagonalization lemma and the corresponding quantification over infinite classes. The existence of $H$ is (metamathematically) *provable* in NAFL, unlike classical and intuitionistic or constructive logics, and so $H$ *really* outputs the correct decision on whether a given legal program halts or not on a legal data input. Hence the contradiction that results from Turing's argument only establishes $X(X)$ as an illegal, self-referential construct in NAFL, rather than the non-existence of $H$.

Infinite sets do not exist in consistent NAFL theories (as established by the detailed formal argument in Sec. 3 of Ref. [7]). The intuitive reason is as follows. An infinite set can only be defined via universal quantifiers, which, in NAFL, refer to the universal class [8]; this makes the above definition self-referential (and hence, illegal in NAFL), since the universal class contains the infinite set in question. Cantor's diagonalization argument is indeed self-referential from the point of view of NAFL in the sense that it requires quantification over (infinitely many) infinite classes, effectively treating these as infinite sets. Gödel's incompleteness theorems and Turing's argument both lead to (or more correctly, *presume*) the existence of non-standard models of Peano arithmetic (PA), a contradiction in NAFL [8]. Indeed, non-standard integers can only be modelled by infinite sets and quantification over these is unavoidable in non-standard models of PA, which therefore cannot exist in NAFL. This fact clearly establishes the infinitary nature of the arguments of Gödel and Turing (and Cantor's diagonal argument), which makes these illegal in NAFL.

The existence of $H$ in NAFL must be formalized in such a manner as to avoid self-reference that leads to quantification over proper classes. Here we run into an apparent difficulty. A classical Turing Machine (TM) is an infinite object in the sense that it has an infinite tape, representing an infinite memory. There are infinitely many non-halting TM's, each of which may use up an infinite amount of memory and the issue is how one may legally quantify over such TM's in NAFL. Here it is crucial to note that the NAFL version of $H$ only formalizes the infinitely many *proofs* that TM's halt or not halt, given their *programs* (which are finite instruction sets, or transition rules), rather than formalize the TM's themselves. In particular, a proof that a program does not halt (or halts) is a finite object and such proofs can be encoded as natural numbers or binary strings and quantified over. Intuitively, one may consider a proof of non-halting as representing a "potential" infinity of computational states assumed by the TM. Hence decidability of the halting problem is essential even for the purpose of formalizing it legally in NAFL. Contrast the situation in FOPL, where the undecidability of halting that results from Turing's argument forces the existence of infinitely many TM's that do not halt with no corresponding proofs of non-halting. In the absence of a proof, one can only represent non-halting as a genuine infinite class of computational states assumed by the TM, which may be thought of as a "completed infinity" and illegal in NAFL. Thus one sees again that Turing's argument forces quantification over infinite classes in order to even formalize the halting problem, thereby effectively requiring the existence of infinite sets.

Exactly the same reasoning as above shows that the algorithm $W$ of Sec. 2 must also exist in NAFL. The master virus $V$ of Sec. 3 that is defined impredicatively, using the class of all viruses, is an illegitimate, self-referential entity (requiring illegal quantification over infinite classes) in NAFL theories. It immediately follows that an autonomic computing system (ACS) as defined in Sec. 2, can indeed exist in NAFL. Note however, that we have not *proven* that the ACS of NAFL must necessarily use *only* classical computing techniques (Turing machines). Indeed, NAFL is a non-classical logic that does permit quantum superposition and entanglement, which are essential ingredients of quantum algorithms, and our proof that $W$ (and $H$) must exist is a metamathematical and non-classical proof that uses the Main Postulate of NAFL. This is elaborated upon in the ensuing section, where we demonstrate that quantum algorithms may indeed be essential for implementing our theoretical model of an ACS in NAFL.

## 7. The link between autonomic and quantum computing in NAFL

As noted in Sec. 2, the various problems (in the infinite class $P$) that afflict given initial states (in the infinite class $S$) of a given autonomic computing system (ACS) must not have access to the functioning of the autonomic manager (AM). Such access would make the purported ACS fail and require human intervention. AM, among other tasks, uses the algorithm $W$ to decide whether a given problem is correctly and optimally handled by a given initial state of the ACS. Indeed, if a problem $P_0$ can access AM, it also has access to $P$, $S$ and $W$, which clearly makes $P_0$ a self-referential entity that is defined using the class of 'all' problems $P$. NAFL (unlike FOPL) bans such self-referential entities, which is the essential reason for the failure of Turing's argument in NAFL (see Sec. 6). At this stage the reader might wonder how one can prevent a determined hacker, who wishes to create $P_0$, from accessing the code implementing AM, and in particular, the algorithm $W$ and those that enumerate $P$ and $S$ (which are used by AM). Any fully classical encryption technique, that does not make use of truly random numbers, can be broken. Hence FOPL offers no foolproof method for preventing the hacker from accessing AM, creating $P_0$ by using the above-mentioned (classical) algorithms, and thereby make the ACS fail and in fact, nonexistent via Turing's argument.

Let us now consider how the above difficulty may possibly be resolved in NAFL. We noted in Sec. 6 that NAFL requires the algorithms $H$ and $W$ to exist and further, there must exist algorithms that enumerate the infinite classes $P$ and $S$; denote these by $P^*$ and $S^*$ respectively. However, the *proof* that we gave in Sec. 6 for the existence of these algorithms (*i.e.*, the proof that any infinite class must be recursive in NAFL) was both *non-classical* and *metamathematical*; this proof used the Main Postulate of NAFL. Indeed, as noted earlier, the Main Postulate uses concepts like undecidability of a proposition within a theory, the consistency of a theory and the existence of a non-classical model for a theory, all of which are strictly metamathematical concepts that reside in the metatheory and cannot be formalized in the language of NAFL theories. These metamathematical concepts cannot be admitted in either the theory syntax or the proof syntax of NAFL theories, essentially because admitting them would amount to illegal self-reference and require a violation of the Main Postulate. A very important non-classical feature of NAFL is that it requires any theory that proves the existence of infinitely many objects satisfying a given property to also prove the existence of the corresponding infinite class of such objects. While NAFL also requires that any such infinite class must be recursive, the algorithm that enumerates this infinite class *need not* be classical and its existence *need not* be formalizable within the theory (that proves the existence of the infinite class in question), because the *proof* of said recursiveness appeals to the Main Postulate of NAFL; see the first paragraph of Sec. 6. However, if existence of the said algorithm cannot be formalized within the NAFL theory, then it must necessarily exist in the metatheory, in which non-classical models for that theory reside. Note that we have accepted a modified Church-Turing thesis in the following form: *Any recursive class must be effectively computable (i.e., there must exist an algorithm that computes it), although not necessarily by classical Turing machines*. Thus "recursive" or "computable" in NAFL does not necessarily mean "computable by a classical Turing machine". Here one must keep in mind that the non-classical features of NAFL mentioned above cause many classical results in recursion theory (due to Gödel and Post) to fail in NAFL. An intuitive explanation for this thesis was given in Sec. 6. We will elaborate further on the necessity for this thesis in future work that deals with real analysis in NAFL. It follows that the algorithms $H$, $W$, $P^*$ and $S^*$ need only exist metamathematically and can be non-classical, essentially because the existence proof for these algorithms appeals to the metamathematical and non-classical concepts inherent in the Main Postulate. It should be emphasized that the formalism (represented by the proof syntax and the theory syntax) of NAFL theories, whose rules of inference are classical, will admit only classical concepts; the Main Postulate further restricts the formally admissible concepts to be non-self-referential.

Let us suppose that NAFL can formalize quantum mechanics as the theory QM. Quantum algorithms will reside in the metatheory of QM, because these use the strictly metamathematical concepts of quantum superposition and entanglement, which are justified using the Main Postulate of NAFL (see Sec. 5). If our model of an ACS can also be formalized in QM, then the solution to the self-reference problem noted earlier is to use quantum algorithms for the autonomic manager AM. In particular, our suggestion is that the algorithms $W$, $P^*$ and $S^*$ (and possibly other algorithms of AM as well) be encoded using quantum computing techniques. Note that we restrict $S$ and $P$ to be classical, *i.e.*, the initial states of the ACS and the problems that afflict them are modelled using classical algorithms. Indeed, only classical algorithms can be formalized and encoded into the theory syntax of NAFL theories, in particular, QM. Clearly, requiring AM to be encoded using quantum algorithms makes it a metamathematical entity, residing in the metatheory of QM, and naturally eliminates the self-reference problem in the sense that no element of $P$ or $S$ can access AM and its algorithms $W$, $P^*$ and $S^*$. Further, to make our proposed scheme practical, AM can be protected using quantum encryption techniques, which, if realized, are *in principle* unbreakable (see Ref. [18] for a recent review article on quantum cryptography). This will ensure that a hacker will never be able to access the code for AM.

In conclusion, NAFL provides the natural logical framework (via syntax and semantics) for the theoretical analysis of autonomic computing systems, including establishment of the link with quantum computing. The use of quantum computing and quantum encryption techniques makes AM a non-classical, metamathematical entity in NAFL that supervises the infinitely many (classical) initial states and (classical) problems of an ACS, represented in the infinite classes $S$ and $P$ respectively. If one can establish the existence of a quantum algorithm for $W$ in NAFL, this would vindicate the work of Kieu [9-12], who argues for a quantum algorithm that solves Hilbert's tenth problem (equivalent to the halting problem). Finally, the reader may object that if quantum computing can be used in AM, why should $S$ and $P$ be restricted to be classical? Since Cantor's diagonal argument fails in NAFL, removing the above restriction should in principle be possible; however, if quantum algorithms are permitted in $S$ and $P$, the entire ACS (rather than just AM) will become a metamathematical entity in NAFL. The model theory for the non-classical models of NAFL theories (in particular, QM) must be developed in a paraconsistent logic [25], in whose theories contradictions are provable. The existence of a fully quantum ACS, that permits quantum algorithms in $S$ and $P$ as well as in AM, can be studied in the framework of such a paraconsistent model theory. Incidentally, note that the Main Postulate of NAFL, which requires the existence of non-classical models for consistent theories with undecidable propositions, actually justifies the need for a paraconsistent logic, something that was not apparent earlier. It should be emphasized that NAFL is *not* a paraconsistent logic in the strict sense because *provability* of contradictions is not permitted in consistent NAFL theories. Indeed, the NAFL notion of consistency, which requires the existence of non-classical models for certain theories as noted above (in addition to the classical models), is actually *stronger* than the corresponding classical notion and severely restricts classical infinitary reasoning.

## 8. Concluding remarks

This paper highlights the importance of logical considerations in deciding whether an autonomic computing system (ACS) can exist in principle. Classical first-order predicate logic (FOPL) makes the prediction that an ACS that uses only standard computing techniques (*i.e.*, classical Turing machines) cannot exist because of Turing noncomputability via Cantor's diagonalization argument. We have shown how Turing's argument fails in NAFL, which therefore does permit an ACS to exist and facilitates the link between autonomic and quantum computing. Future work will outline how a limited version of real analysis can be executed in NAFL, despite the non-existence of infinite sets and the illegality of quantification over infinite proper classes within NAFL. This should pave the way for an axiomatization of quantum mechanics in a single logic (NAFL), without the present need to jump from quantum logic to FOPL. Such an axiomatization in NAFL, apart from having profound implications for the foundations of physics, will be very useful for the theoretical study of quantum and autonomic computing systems.

We close with a few brief remarks on quantum logics [19-22] and intuitionistic/constructive logics [23, 24], and contrast them with NAFL. Quantum logics have been developed as *empirical* tools to handle certain "weird" aspects of quantum mechanics, such as, superposition and entanglement; see Sec. 17 of Ref. [19], where the authors remark:

"It seems to us that quantum logics are not to be regarded as a kind of 'clue', capable of solving the main physical and epistemological difficulties of QT [quantum theory]. This was perhaps an illusion of some pioneering workers in quantum logic."

The empirical and *ad-hoc* nature of quantum logic means that it cannot be used to justify abstract aspects of quantum mechanics which require, for example, real analysis, where FOPL is used. Therefore, presently one has to abruptly jump from quantum logic to FOPL (*i.e.*, develop a kind of logical "schizophrenia", as noted by M. L. Dalla Chiara) in order to understand quantum mechanics. The absence of a single logical framework for quantum mechanics makes it difficult, if not impossible, to resolve outstanding issues like the (in)compatibility between relativity theory and quantum mechanics. Further, conceptual treatment of subjects like autonomic computing and computability theory is not possible in quantum logic. In contrast, NAFL justifies quantum superposition and entanglement via a deep metamathematical principle (the Main Postulate) which is universally applicable in all contexts. Hence NAFL, if developed to its full potential, can provide a single logical framework in which quantum mechanics as well as quantum and autonomic computing systems can be studied. The axiomatic nature of NAFL truth implied by the Main Postulate has very deep implications for the foundations of theoretical science and mathematics, and deserves to be studied further.

Intuitionistic and constructive logics [23, 24] also deny the law of the excluded middle and attempts have been made to justify certain aspects of theoretical physics, including quantum mechanics, in these logics. These logics nevertheless uphold the law of non-contradiction, which is very problematic [6] from the point of view of internal consistency; see the last paragraph of Appendix A for a critique of the intuitionistic law of non-contradiction. As a consequence, intuitionistic and constructive logics admit infinite sets and Cantor's diagonalization argument. The resulting undecidability of the halting problem means that, in particular, truly autonomic computing systems cannot exist in these logics, as opposed to NAFL.

# Appendix A. Proposition 1 and its proof

Here we reproduce Proposition 1 and its proof from Ref. [6].

**Proposition 1.** *Let Q be an undecidable proposition in a consistent* NAFL *theory* T. *Then* $Q \vee \neg Q$ *and* $\neg(Q \& \neg Q)$ *are not theorems of* T. *There must exist a non-classical model for* T *in which* $Q \& \neg Q$ *is the case*.

*Proof.* By the Main Postulate of NAFL, $Q$ ($\neg Q$) can be the case in T if and only if $Q$ ($\neg Q$) has been asserted *axiomatically*, by virtue of its provability in T*. In the absence of any such axiomatic assertions, (*e.g.* if T* = T), it follows that neither $Q$ nor $\neg Q$ can be the case in T and hence $Q \vee \neg Q$ cannot be a theorem of T. The classical refutation of $Q \& \neg Q$ in T proceeds as follows: 'If $Q$ ($\neg Q$) is the case, then $\neg Q$ ($Q$) cannot be the case', or equivalently, '$\neg Q$ ($Q$) contradicts $Q$ ($\neg Q$)'. But by the Main Postulate, this argument fails in NAFL and amounts to a refutation of $Q \& \neg Q$ in T* = T + $Q$ (T + $\neg Q$), and *not* in T as required. Careful thought will show that the classical refutation of $Q \& \neg Q$ in T is the *only possible* reason for $\neg(Q \& \neg Q)$ to be a theorem of T, and it fails in NAFL. The intuitionistic (constructivist) refutation of $Q \& \neg Q$ in T is flawed and also fails in NAFL, as will be shown in the ensuing paragraph. By the completeness theorem of FOPL, which is taken for granted in NAFL, it follows that there must exist a non-classical model for T in which $Q \& \neg Q$ is satisfiable. □

Consider the law of non-contradiction as stated in a standard system of intuitionistic logic due to S. C. Kleene, namely, $\neg Q \Rightarrow (Q \Rightarrow R)$. This formula asserts that from contradictory premises $Q$ and $\neg Q$, an *arbitrary* proposition $R$ can be deduced, which is absurd. Hence $\neg(Q \& \neg Q)$ seemingly follows. However, note that in intuitionism, truth is provability (not necessarily in a specific theory T); together with the intuitionistic concept of negation, it follows that an assertion of $\neg(Q \& \neg Q)$ is the same as deducing an absurdity from $Q \& \neg Q$, or equivalently, from contradictory premises $Q$ and $\neg Q$. But we have seen that the 'absurdity' referred to here is precisely the fact that *any proposition can be deduced*, given contradictory premises! The above 'proof' of $\neg(Q \& \neg Q)$ from contradictory premises, mandated by the intuitionistic concepts of truth and negation, is flawed because *any proposition can be so deduced*. Note that this 'proof' is formally indistinguishable from one in which $\neg(Q \& \neg Q)$ is *substituted* for the deduced arbitrary proposition $R$. In NAFL, it is not possible to deduce an arbitrary proposition from contradictory premises [6, 7], and so the flawed intuitionistic argument for $\neg(Q \& \neg Q)$ fails in any case.

## Appendix B. Failure of relativistic locality in NAFL

Let X and Y be distant, stationary particles in the inertial frame of reference F1. Let *A* and *B* be probabilistic (undecidable-in-QM) spacetime events related to X and Y respectively, such that quantum entanglement requires $A \Leftrightarrow B$ to be in the proof syntax of QM in the F1 frame. For example, *A* could express that

"The *y*-spin of X at the coordinates $(x, t) = (0, 0)$ is +1/2",

while *B* stands for

"The *y*-spin of Y at the coordinates $(x, t) = (\xi, 0)$ is –1/2".

Let F2 be another inertial frame moving relatively to F1 as shown in Fig. 2, where F1-O and F2-O are observers at the origins of the respective frames and coinciding with the particle X at $(t, t') = (0, 0)$.

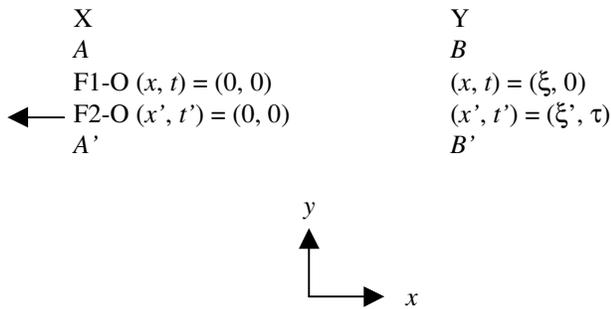

Figure 2. Entanglement in the relativistic scenario

Let F1-O observe *A*, and hence, deduce *B* by entanglement, at $t = 0$. Further, no measurements are made prior to $t = 0$, so that the superpositions $A \& \neg A$ and $B \& \neg B$ hold for F1-O at times $t < 0$. Similarly, F2-O observes *A'* (which is just *A* expressed in the indicated F2 coordinates) at $t' = 0$ and the superposition $A' \& \neg A'$ holds for F2-O at times $t' < 0$. By the Lorentz transformations, F2-O (at time $t' = 0$) deduces *B'*, which is *B* expressed in the F2 coordinates indicated in Fig. 2, with $\tau > 0$. This is so because the Lorentz transformations require $A' \Leftrightarrow B'$ to be a theorem in the F2 frame. But for $t' < \tau$, the Main Postulate of NAFL requires the superposition $B' \& \neg B'$ to hold for F2-O. In particular, for $0 < t' < \tau$, F2-O deduces $A' \& \neg B'$, in violation of the theorem $A' \Leftrightarrow B'$ required by the Lorentz transformations. This is essentially the NAFL objection to relativistic locality and the Lorentz transformations [13-16]. Indeed, for $0 < t' < \tau$, the probabilistic (undecidable-in-QM) spacetime event *B'* has not yet occurred for F2-O, and so F2-O could in principle axiomatically assert either *B'* or even ¬*B'*, or choose not to make either assertion (in which case the superposition $B' \& \neg B'$ holds), by the Main Postulate of NAFL. But special relativity theory does not recognize time as an explicit entity with the division of events into present, future and past, and consequently, requires F2-O to assert *B'* at $t' = 0$, thereby "instantaneously collapsing" the superposed state of $B' \& \neg B'$ to *B'*. Thus special relativity theory does not permit the possibility ¬*B'* (and hence, even the superposed state $B' \& \neg B'$) for F2-O at times $0 < t' < \tau$, in violation of the Main Postulate of NAFL. Note that *by definition* of a probabilistic event, *B'* must be undecided for F2-O at times $t' < \tau$, leaving open either possibility *B'* or ¬*B'*. The argument given above shows that special relativity theory is intolerant of intrinsically undecidable, probabilistic events even within classical logic [14-16]. Note also that prior to $t = 0$ ($t' = 0$) both *A* and *B* (*A'* and *B'*) are unambiguously in the superposed state for F1-O (F2-O), by the Main Postulate of NAFL. In particular, $A \& \neg B$ ($A' \& \neg B'$) hold for F1-O (F2-O) prior to $t = 0$ ($t' = 0$), and the Lorentz transformations are again violated.